\documentstyle[twocolumn,epsfig,aps]{revtex}

\input{psfig.sty}

\begin{document}
\draft
\title{A storage ring for neutral atoms}
\author{J. A. Sauer, M. D. Barrett, and M. S. Chapman}
\address{School of Physics, Georgia Institute of Technology, \\
Atlanta, Georgia 30332-0430}
\maketitle

\begin{abstract}
We have demonstrated a storage ring for ultra-cold neutral atoms. Atoms with
mean velocities of 1 m/s corresponding to kinetic energies of $\sim $100 neV
are confined to a 2 cm diameter ring by magnetic forces produced by two
current-carrying wires. Up to 10$^{6}$ atoms are loaded at a time in the
ring, and 7 revolutions are clearly observed. Additionally, we have
demonstrated multiple loading of the ring and deterministic manipulation of
the longitudinal velocity distribution of the atoms using applied laser
pulses. Applications of this ring include large area atom interferometers
and cw monochromatic atomic beam generation.
\end{abstract}



\narrowtext

Controlling the motional degrees of freedom of neutral atoms has emerged as
a major theme in modern atomic physics. The development of laser cooling and
related techniques to cool atoms down to the $\mu $K regime allows trapping
and manipulation of the atomic motion using interactions with the relatively
weak electric and magnetic fields available in the laboratory. Complementing
the 3-D trapping work, there has been considerable progress recently in
developing 2-D magnetic guiding structures for ultra-cold neutral
atoms---so-called guided atom optics. Much of this work is motivated by the
prospects of coherent de Broglie wave transport for applications including
ultrasensitive atomic interferometers and quantum information processing. \
A variety of simple guides have been developed based on current-carrying
wires\cite{ASchmiedemeyer,BCornell,CPrentiss,4wiretube,raithel4wire}, and
extensions of these systems have been used to demonstrate atomic
beamsplitters\cite{Schmiedesplitter,Dandersonsplitter}, atom conveyors\cite
{Hanschconveyor}, and magnetic microtraps \cite{Hanschconveyor,Schmiedechip}.

It is compelling to extend these techniques to a ring geometry, and there
have been several proposals along these lines \cite
{Proppritchard,PropThompson,Propkatz}, as well as a recent demonstration
confining fast polar molecules to an electrostatic ring \cite{molecuring}. \
These are motivated in part by the prospects of ring-based atomic
interferometry and monochromatic atomic beam generation. \ Similar to
optical ring interferometers, the ring geometry offers greater sensitivity
through increasing the effective enclosed area of the interferometer both by
straightforward enlargement of the ring as well as by employing
multiple-orbit interfering trajectories. This is important because the
sensitivity of many interactions (e.g. Sagnac rotational phase) is
proportional to the enclosed area of the interferometer\cite{atominter}.
This geometry also provides new opportunities to create ultra-cold and
monochromatic atomic beams. For example, the ring can be multiply loaded to
increase the number of atoms in the ring, and both the longitudinal and
transverse velocity distributions of the atoms can be manipulated and cooled
in the ring. Finally, highly directional output beams can be created by
using a coherent, variable output coupler.

In this work, we demonstrate the first storage ring for neutral atoms. Our
ring consists of a particularly simple 2-wire magnetic guiding structure. We
have developed a technique to successfully transfer atoms to the ring from
an external magnetic waveguide directly loaded from a magneto-optic trap
(MOT). We observe up to 7 complete revolutions of the atoms in the ring, and
we have also demonstrated multiple loading. Finally, we have performed
simple manipulations of the longitudinal velocity distribution of the atoms
in the ring.

Magnetic confinement of neutral atoms is based on the interaction of the
atomic magnetic dipole projection $\mu _{m}$ with a spatially varying
external magnetic field ${\bf B}({\bf r}).$ For slowly moving atoms, the
dipole will adiabatically follow the field direction, yielding a trapping
potential energy $U({\bf r})=-\mu _{m}\left| {\bf B}({\bf r})\right| $.
Hence, atomic states with spins anti-aligned with the local field ({\em i.e.}
with $\mu _{m}<0,$ so called weak-field seeking states) will be confined to
regions in space with a minimum in the magnetic field magnitude.

To provide this minimum, we use a 2-D quadrupole magnetic field produced by
two nearly parallel wires carrying equal currents in the same direction. To
lowest order, the field between the wires is given by ${\bf B}(x,y)=4\mu
_{0}I(y\hat{x}+x\hat{y})(\pi d^{2})^{-1}$ \ where $\mu _{0}$ is the usual
permittivity of free space, $d$ is the spacing between the wires, and $I$\
is the current in each wire. A plot of the full trapping potential (ignoring
gravity) is shown in Fig 1c; the corresponding trap depth is $I\mu _{0}\mu
_{m}/d\pi ,$ about 1/2 that of the more common 4-wire guide.

Our experiment utilizes 2 pairs of such wires as shown in Fig. 1a---the
``ring'' that confines the atoms to a circular path, and the ``guide'' that
couples atoms from the MOT\ into the ring. Both the guide and the ring
consist of 280 $\mu $m diameter copper wires capable of sustaining a steady
state current of several amps. \ \ The ring wires have a separation $d\sim $
840 $\mu $m, which, at a wire current of 8 amps, provides a field gradient
of 1800 G/cm, a mean trap frequency of 590 Hz, and a trap depth of 2.5 mK
for the $F=1,$ $m_{F}=-1$ ground state of $^{87}$Rb. The diameter of the
storage ring is 2 cm. The spacing of the guide wires varies from 4 mm where
the MOT is formed down to 840 $\mu $m where it overlaps with the ring. \ \

We begin the experiment by loading $^{87}$Rb atoms into a MOT between the
guide wires. The MOT\ is produced by 3 retro-reflected laser beams, each
having an intensity of 12 mW/cm$^{2}$ and a $1/e^{2}$ diameter of 1 cm. The
trap lasers are tuned 17\ MHz below the $5S_{1/2}-5P_{3/2}$ $F=2$ to $%
F^{\prime }=3$ transition, while anti-Helmholtz coils generate a magnetic
field gradient of 6 G/cm. An additional laser beam tuned to the $F=1$ to $%
F^{\prime }=2$ transition with an intensity of 4 mW/cm$^{2}$ repumps the
atoms decaying into the $F=1$ state. The MOT is loaded directly from a
thermal beam, and the trap typically contains $\sim $6 x $10^{6}$ atoms
after 2 s of loading. The presence of the guide wires directly in the path
of the trap laser beams necessitates careful alignment of the MOT beams and
coils---perhaps because the shadows of the wires reduce the effective
loading volume of the MOT. Following loading, the MOT coils are turned off
and the guide current is ramped on in 5 ms. A short interval of sub-Doppler
cooling is performed, during which the trap laser detuning is ramped further
to the red by 110 MHz over 2 ms and the repump intensity is lowered ten-fold
and finally shuttered off. At this point the atoms are all in the $F=1$
ground state with a measured longitudinal temperature 3 $\mu $K. \ From the
extent of the cloud, we infer a transverse temperature of 57 $\mu $K after
the guide current reaches its final value.

About 15\%\ of the atoms in the MOT\ are transferred to the guide; this is
comparable to other experiments using this technique \cite
{4wiretube,raithel4wire}. We have found that the coupling efficiency
increases with increasing guide current, however\ our present set-up is
limited to a maximum current of 8 amps. \ Once in the guide, the atoms fall
4 cm under gravity to the 15 mm overlap with the ring. To transfer the atoms
from the guide to the ring, we ramp the current in the guide off while
simultaneously increasing the current in the ring to its final value. \ This
process transfers the trap center from the guide to the ring (see Fig. 1b).
\ Optimal transfer is achieved for equal currents in the guide and ring. \
The transfer efficiency to the ring is estimated to be
\mbox{$>$}%
90\% and is maximized by a transfer time of 16 ms. \ For longer transfer
times, the cloud traverses the entire overlap region before transferring
completely to the guide---for shorter times, we measure losses from the
cloud, possibly due to heating. \ \

To measure the evolution of the atoms in the ring, they are resonantly
excited with a 1 ms laser pulse focused directly between the wires of the
ring, and their florescence is imaged onto an intensified CCD camera (see
Fig. 2). \ By repeating the measurement with different probe delay times,
the complete trajectory of the atoms can be measured. A typical measurement
of the atomic orbits in the ring is shown in Fig. 3. The different peaks
correspond to complete revolutions of the ring, and 7 complete revolutions
of the ring are clearly distinguished. The measured orbit time of 81 ms
corresponds to an average velocity of 85 cm/s, which is consistent with a 4
cm free-fall. The peaks fit reasonably well to a simple model incorporating
only losses from the ring as well as the continued azimuthal free expansion
of the atom cloud. From this fit, the 1/$e$ lifetime of the ring is
determined to be 180 ms, and the azimuthal temperature is measured to be \ $%
3.4(3)$ $\mu $K$,$ only slightly hotter than the temperature measured
immediately after loading the guide. To obtain more satisfactory agreement
with the data, the model was extended to include a term corresponding to a
loss of atoms from the orbit to a diffuse background in the ring.

The lifetime of the ring is somewhat shorter than expected from losses due
to background collisions alone. Although we do not have a reliable direct
measurement of the vacuum at the ring, we infer a vacuum-limited lifetime of
\mbox{$>$}%
$800$ ms from the MOT loading time constant. Additionally, we have measured
that the lifetime of the storage ring decreases by 20\% if the current in
the ring is ramped from 8 to 5 amps over $40$ ms after the atoms are loaded.
Possible additional loss mechanisms from the ring include non-adiabatic
spin-flips (so-called Majorana transitions) and non-ergodic mixing from the
azimuthal motion to the transverse motion. Adiabatic following of the
magnetic field requires the time-rate of change of the field direction to
vary slowly compared with the precession frequency, $d\theta /dt<\mu
_{m}|B|/\hbar $\cite{ketterlereview} ---where the field goes to zero, this
condition cannot be met and hence atoms passing within a minimum radius of
the guide center are likely to be lost from the ring . Following the model
given in \cite{TOPpaper}, the loss radius for our guide is $b_{0}=0.6$ $\mu $%
m, which, together with the cloud size and transverse temperature, leads to
an expected 1/$e$ lifetime of 300(100) ms . \ However, we would expect that
the loss rate would increase with increasing ring current, which is contrary
to our observations. In any case, this issue can be readily avoided in
future rings by adding a single axial wire to the ring, which would provide
a small azimuthal field.

The losses could also be the result of non-ergodic mixing due to
imperfections in the ring. \ A likely suspect in this regard is the junction
in the ring where the current is fed in and out. \ We estimate that the
non-uniformity in the current distribution at this location produces a 20\%
ripple in the potential over a 250 $\mu $m distance. This perturbation in
the magnetic potential could transfer some of an atom's azimuthal energy
into transverse energy and cause it to leave the trap. \ From the extent of
the cloud inside the ring, we infer a transverse temperature of 1.35 mK
after one revolution, about 2 times greater than expected from adiabatic
heating alone. The average orbital kinetic energy of the atoms is about
twice the trap depth at 8 amps, and hence an atom need only transfer 15\% of
its azimuthal energy to transverse energy to be lost from the trap in two
revolutions. \ This mixing could also explain the slowly changing diffuse
background mentioned above. Furthermore, we would expect that this loss
mechanism would decrease with increasing current as the trap becomes deeper,
while the atoms are compressed to the center of the guide where the
perturbation is weakest. In future rings, the field irregularity can be
greatly minimized with additional trim wires.

Even with this ring lifetime, it is noteworthy that the total guided
distance in the ring is $\sim $0.5 m and if this system was used in a ring
interferometer configuration, the enclosed area would be $\sim $4400 mm$^{2}$%
, 200-fold larger than the most sensitive atomic gyroscope
demonstrated to-date \cite{KasevGyro}. Furthermore, with
straightforward modifications to the ring discussed above,
together with improved vacuum conditions, it should be possible to
increase these values by another 100-fold
(corresponding to a ring lifetime 20 s), yielding an intrinsic sensitivity 10%
$^{4}$ larger than in \cite{KasevGyro}. Of course maintaining atomic
coherence over these time scales is a significant technical challenge, and
would most certainly require guiding in a single transverse mode, however
the ring geometry offers important advantages. \ In particular, by using
complete revolutions of the ring for the interfering counterpropagating
trajectories, we can ensure that each interfering amplitude acquires the
same dynamical phase from the guiding potential (to the extent than the ring
current and dimension remain constant), which will be important to cancel
out effects of inevitable irregularities in the guiding potentials. This
advantage, together with the recent demonstrations of fast, compact
techniques for making atomic Bose condensates in optical \cite{opticalBEC}
and magnetic microtraps \cite{MICROBEC} makes the prospects of single-mode
guided atom interferometry quite promising.

The ring geometry also provides unique capabilities for the generation of
intense, cold atomic beams. Although linear guides are already promising in
this regard \cite{Dalibard}, the ring geometry allows for convenient
multiple or even continuous loading. Furthermore, the ring could be combined
with a very selective output coupling mechanism (employing laser induced
Raman transitions for example) to provide a very bright extracted beam.
Here, the ring configuration can provide efficient `recycling' of the
unextracted atoms, which can then be further manipulated in the subsequent
revolutions. We have taken first steps toward these goals by demonstrating
multiple loading of the ring and simple manipulation of the azimuthal atomic
momentum distributions.

To multiply load the ring, we immediately begin to reload the MOT after the
ring is initially loaded. After a 0.2 s loading time, these atoms are loaded
in the guide and fall down to the ring. \ To load them into the ring without
releasing the originally loaded atoms, the ring current is first ramped down
to 2 A to allow the reloaded atoms into the overlap region. Then the ring
current is then increased while the guide current is decreased just as in
the original transfer. \ Fig. 4 depicts a typical double loading trajectory.
The additional loading of the ring appears as a second set of peaks orbiting
$180^{\circ }$ out of phase with the original loading. \ Formation of the
second MOT\ causes some losses in the original cloud due to scattered light
and varying magnetic fields, however these losses can be avoided with
appropriate isolation of the ring. Although the technique we used for this
demonstration is limited to separated pulses, it is also possible to design
potentials that allow continuous loading. Such potentials are necessarily
`bumpy' (otherwise it would be possible to increase phase space density,
which would violate Louiville's theorem), however the additional entropy
could be removed by subsequent laser cooling or evaporative cooling in the
ring.

In our final experiment, we used a resonant beam to alter the azimuthal
velocity distribution of orbiting atoms. The original azimuthal velocity
distribution has a width $\Delta v\sim 2$ cm/s corresponding to a speed
ratio $v/\Delta v\sim 50$ in the ring. Because of the free expansion of the
atomic cloud along the guide direction, the initial velocity distribution is
mapped onto the azimuthal spatial distribution of the atoms in the ring.
Hence the velocity distribution can be modified by selective temporal or
spatial removal of portions of the cloud. \ Two such modifications are shown
in Fig. 5. In both cases, the velocity distribution has been modified in the
previous orbit (not shown) and traces show the evolution for subsequent
orbits. In the bottom trace, only the central 40\% of the FWHM has been
preserved, corresponding to an increase of the speed ratio to 125. In the
top trace, the central 40\% of the FWHM has been removed, leaving a
double-peaked distribution. In both cases, the subsequent evolution and
broadening of the peaks occur at the expected rate. \

In summary, we have demonstrated a magnetic storage ring for neutral atoms,
together with an efficient loading method. We have also demonstrated that
the ring can be multiply loaded, and we have manipulated the velocity
distribution of the atoms in the ring. Extensions of this work to ring-based
atom interferometry and cold-beam generation hold much promise for the
future. \qquad

We would like to acknowledge the technical assistance of D. Zhu and helpful
discussions with T.A.B. Kennedy, C. Raman, and L. You. This work was
partially supported by the National Security Agency (NSA) and Advanced
Research and Development Activity (ARDA) under Army Research Office (ARO)
contract number DAA55-98-1-0370.

\begin{figure}[tbp]
\centerline{\epsfig{figure=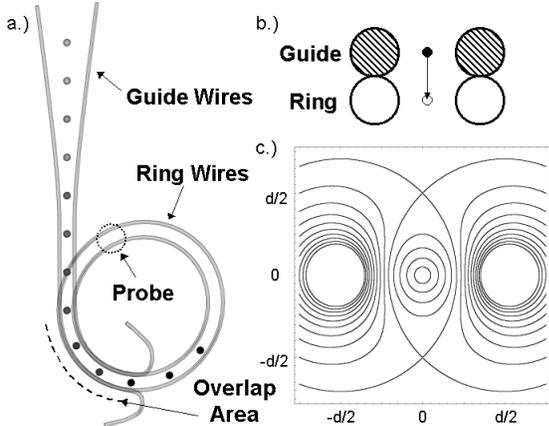,width=3.0in}} \caption{Fig 1.
a) A\ schematic of the storage ring. \ b) \ A cross section of the
overlap region. The trap minimum is shifted from between the guide
wires to the ring wires by adjusting the current. \ c) \ A\
contour plot of a two wire potential. \ The contours are drawn
every 0.5 mK for $d$ = 0.84 mm and $I$ = 8 amps. . \ }
\label{fig1}
\end{figure}

\begin{figure}[tbp]
\centerline{\epsfig{figure=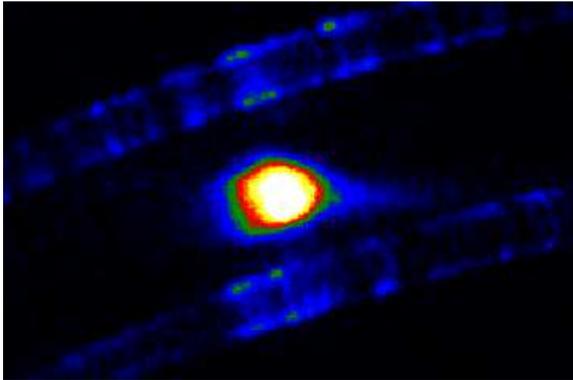,width=3.0in}} \caption{A false
color image shows probing of the atomic cloud in the storage ring.
\ These atoms have made 2 complete revolutions in the ring. \ The
ring wires are visible above and below the cloud, and have a
center to center spacing of 840 $\protect\mu $M.}
\label{4wiretube}
\end{figure}

\begin{figure}[tbp]
\centerline{\epsfig{figure=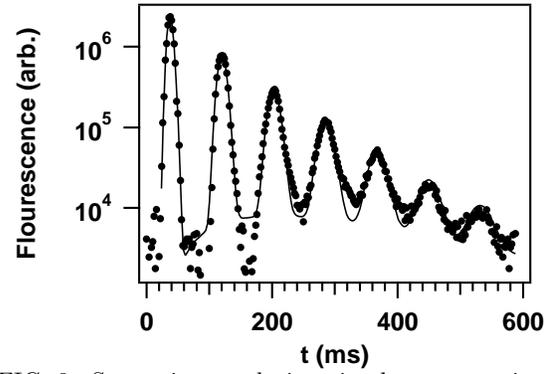,width=3.0in}}
\caption{Successive revolutions in the storage ring. \ The points
represent experimental data, the curve is a theoretical model. \
The first peak corresponds to the first complete revolution in the
ring.} \label{fig3}
\end{figure}

\begin{figure}[tbp]
\centerline{\epsfig{figure=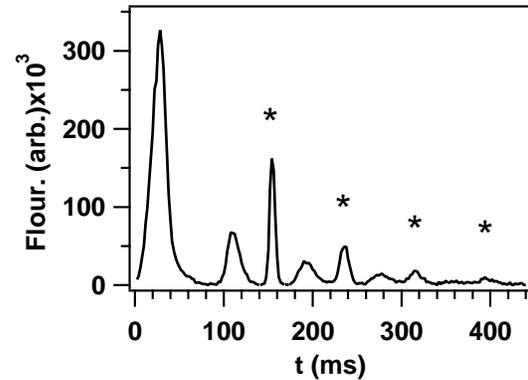,width=3.0in}} \caption{Double
loading into the ring. \ Those peaks marked with a (*)\ are from
the second loading. \ The clouds are 180$^{0}$ out of phase, with
a spacing of only 40.5 ms between successive peaks.} \label{fig4}
\end{figure}

\begin{figure}[tbp]
\centerline{\epsfig{figure=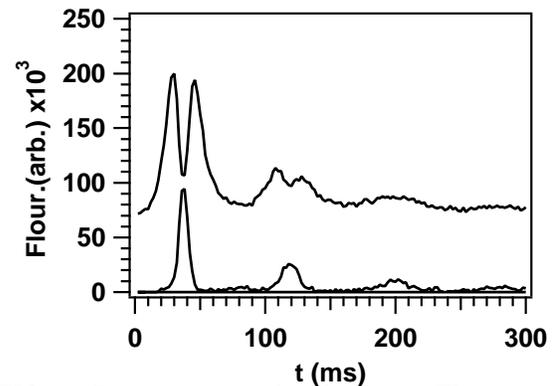,width=3.0in}}
\caption{Deterministic pulse shaping. \ The upper graph
corresponds to a cloud with its center removed. \ The lower graph
has everything but its center removed. \ The upper graph has been
offset for clarity. \ \ } \label{fig5}
\end{figure}

\end{document}